\documentclass{mn2e}

\usepackage{epsf}

\title[Accretion Discs and the Solar Nebula]
{The Accretion Disc Dynamo in the Solar Nebula}

\author[A. R. King and J. E. Pringle]
{A. R. King$^1$ and
J.E. Pringle$^{1,2}$\\ $^1$Theoretical Astrophysics Group, University
of Leicester, Leicester LE1 7RH\\ $^2$Institute of Astronomy,
University of Cambridge, Madingley Road, Cambridge CB3 0HA}

\date{\today}

\volume{000}

\pagerange{\pageref{firstpage}--\pageref{lastpage}} \pubyear{2006}

\begin{document}

\label{firstpage}

\maketitle

\begin{abstract}

The nearest accretion disc to us in space if not time was the
protosolar nebula. Remnants of this nebula thus potentially offer
unique insight into how discs work. In particular the existence of
chondrules, which must have formed in the disc as small molten
droplets, requires strong and intermittent heating of disc
material. We argue that this places important constraints on the way
gravitational energy is released in accretion discs, which are not met
by current shearing--box simulations of MRI--driven dynamos. A deeper
understanding of accretion energy release in discs may require a
better model for these dynamos.

\end{abstract}


\section{Introduction}

It is now generally agreed that accretion discs are driven mainly by
magnetic torques (Shakura \& Sunyaev, 1973) and that the magnetic
fields in such discs are maintained by local dynamo processes. What is
not understood however, despite considerable theoretical efforts, is
how such dynamo processes work, and exactly where and in what form the
accretion energy is released (King et al., 2007; Blackman, 2010).

Given the difficulty of this problem, we should look for observational
evidence which bears on it. We have unique insight into the remains of
one accretion disc -- the protosolar nebula. It is therefore worth
asking if the present--day solar system offers evidence which can
constrain the way energy is released in accretion discs.

Among the oldest solid objects in the solar system are chondrules, the
round grains present in the majority of meteorites. These must have
formed as molten droplets in space before being accreted. The need to
heat them sufficiently implies a connection with energy release in the
protosolar nebula. We consider here the overall energetics required of
the heating process. We argue that since most of the meteoritic
material has been subject to this heating, the energy source required
must be quite widespread and substantial, and must therefore involve
the major local source of energy, i.e. disc accretion.  We find that
around 10 per cent of the accretion energy is required. If so, it is
evident that the existence of chondrules provides fundamental
information about energy release in accretion discs.

This paper is organised as follows. In Section 2 we give a brief
introduction to chondrules and ideas about their formation. In Section
3 we consider the global energetics of the heating process.  A likely
mechanism for the provision of transient heating within the solar
nebula involves strong shocks (Deschl \& Connolly, 2002, Connolly et
al., 2006). In Section~4, we briefly present the model of chondrule
formation through shock heating suggested by Deschl \& Connolly (2002)
but argue that their proposed mechanism for generating these shocks by
gravitational instabilities within the disc requires disc properties
which do not sit easily with our current understanding of disc
evolution. In Section~5 we describe the disc properties which appear
likely to hold at the time of chondrule formation. The need for
significant disc dissipation in the form of shocks, together with
evidence for magnetic fields within the nebula at that time
(Section~6) lead us to a picture of a disc dynamo (Section 7). This
differs markedly from the results of current numerical
simulations. Our picture draws on the dynamo model of Baggaley et
al. (2009a, b) and involves thin flux ropes and reconnection,
analogous to models of flaring on the solar surface.  Section~8 is a
discussion.

\section{Chondrules}

Most meteorites are chondrites (more than 75 per cent, Sears \& Dodd,
1988; Hutchinson 2004), and the most abundant constituent of the
majority of chondritic meteorite groups are chondrules (Grossman et
al., 1988). Chondrules are small particles of silicate material
(typically around a millimetre in size, Weisberg et al., 2006) that
experienced melting before incorporation into chondritic meteorite
parent bodies (Hewins, 1996). The properties of chondrules are
consistent with formation in the protosolar nebula from primitive
material, and thus provide an important record of events in this
disc. Isotopic dating of chondrules indicates ages of 4.5 Gyr, and
also suggests that high temperature nebular processes, such as
chondrule formation, lasted for about $3 - 5$~Myr (Russell et al.,
2006; Scott, 2007). This timescale coincides with the observed
lifetime of protostellar discs around solar--type stars (e.g. Hernandez et
al., 2009). Chondrules formed a few Myr after the earliest CAIs
(Ca--Al--rich inclusions) (Swindle et al., 1996). These formed at
lower temperatures ($\sim 1500$~K) and in general have longer cooling
times (Davis \& MacPherson, 1996), and are part of the same sequence
of heating events.

The source of most meteorites is now recognised as the asteroid belt
(Hutchinson, 2004). The present asteroid belt is thought to consist of
a tiny fraction of the planetesimals that were originally at these
radii, the majority being expelled by dynamical resonances once the
large planets (Jupiter and Saturn) had acquired sufficient mass
(Morbidelli et al., 2005). Thus we may regard the chondritic
meteorites as a trace sample of what remained in the protoplanetary
disc at radii around 3 AU once disc evolution had finished, much of
the solid material had been accreted into planets, and any gas not
accreted to form the giant planets had been expelled. Unless the
dynamical processes which expelled the bulk of the solid material from
the region of the asteroid belt had some means of preferentially
leaving behind chondritic material, it follows that the chondritic
meteorites give some idea of the planetesimals formed in that region,
and therefore of the physical processes in the disc towards the end of
the major accretion phase. Thus, although there are models of
chondrule formation which involve recycling of material radially in
the disc, either by diffusive processes (e.g. Morfill 1983, Clarke \&
Pringle, 1988) or, more plausibly, by a more direct route (Shu et al.,
2001; see also King, Pringle \& Livio, 2007), we shall assume here
that chondrule formation took place more or less {\it in situ} and ask
what implications this might have for the workings of an accretion
disc.

Many different mechanisms have been put forward for the formation of
chondrules (see, for example, the review by Boss, 1996). As discussed
there, the main problem with the formation of chondrules {\it in situ}
is that the temperatures needed in their formation (1500 -- 2000 K,
Hewins, 1988; Hutchinson, 2004; Connolly et al., 2006) greatly exceed
the likely ambient temperatures in the protoplanetary disc at late
times. Higher disc temperatures are possible at earlier times when the
disc was more massive, disc evolution more rapid, and the accretion
rates large, but chondrules formed at that time would mostly have been
carried inwards by the accretion disc and deposited into the Sun.
Chondrule formation in a protostellar disc as it reaches its later
stages of evolution requires a series of transient heating events
within the disc (Connolly et al., 2006). Thus

`Although a very large number of mechanisms have been proposed for making
chondrules, only one has been shown to satisfy most of the constraints:
shock heating of nebula gas and entrained silicates.' (Scott, 2007). 

However the same author also remarks that

`There is no observational evidence for shocks in protoplanetary
disks, and a single source of shocks that could have operated over
several million years has not been demonstrated.' (Scott, 2007).

We attempt this in the present paper.

\section{Energetics of the formation process}

As remarked above, chondritic meteorites appear to come from the
asteroid belt. Accordingly we assume that {\it in situ} chondrule
formation occurs in a region of the protostellar disc at radii in the
range 2 -- 5 AU, and so at around a typical radius of
\begin{equation}
R = 3 \, {\rm AU} = 4.5 \times 10^{13} \, {\rm cm}.
\end{equation}
At this radius disc material has circular velocity
\begin{equation}
  V_\phi = \left( \frac{GM}{R} \right)^{1/2} = 1.7 \times 10^6
  \left( \frac{M}{M_\odot} \right)^{1/2} \left( \frac{R}{ 3\, {\rm AU}} \right)
^{-1/2}\, {\rm  cm \, s}^{-1},
\end{equation}
and thus a specific kinetic energy
\begin{equation}
  E_{\rm kin} = 
\frac{1}{2} V_\phi^2 = 1.5 \times 10^{12} 
  \left( \frac{M}{M_\odot} \right) \left( \frac{R}{ 3\, {\rm AU}} \right) \,
{\rm erg \, g}^{-1}.
\end{equation}

We can think of this as being equivalent to the accretion energy
available locally. This is the energy source that we shall need to use
to power chondrule formation.

The specific energy required to heat the material from a few hundred K
up to $\sim 1800$~K and then melt it is given by Wasson (1996) as
\begin{equation}
E_{\rm melt} = 2.1 \times 10^{10} \, {\rm erg \, g}^{-1}.
\end{equation}
To form chondrules it is necessary to heat the proto--chondrules (or
`dust balls') to melting point. The fraction of chondrule to matrix
(the remainder of the chondrite, which has never melted) in chondrites
is large, and in addition there is evidence that about 15 per cent of
chondrules have undergone more than one melting event. Thus we shall
assume that {\it on average} each proto-chondrule (or `dust-ball') is
subject to around one melting event.

Since $E_{\rm kin} \gg E_{\rm melt}$, it seems there is plenty of
energy available. However, this is not the whole story. Once heated
to melting point, the chondrules must cool. If they simply radiated
their energy into space as black bodies they would cool on a timescale
\begin{equation}
t_{\rm rad} = \frac{E_{\rm melt} m}{4 \pi r^2 \sigma T^4},
\end{equation}
where a chondrule is assumed to have mass $m$, radius $r$ and
temperature $T$, and $\sigma$ is the Stefan-Boltzmann constant.  Thus
a typical cooling time for a chondrule would be (see, for example, the
cooling curves given by Wasson, 1996)
\begin{equation}
 t_{\rm rad} = 3.5 \left( \frac{E_{\rm melt}}{2.1 \times 10^{10} {\rm erg/g}} 
\right) \left( \frac{r}{1 {\rm mm}} \right) \left(
  \frac{\rho_{\rm chon}}{ 3 {\rm g/cm}^3} \right) \left( \frac{T}{1800 \, {\rm
      K}} \right)^{-4} \, {\rm s},
\end{equation}
where $\rho_{\rm chon}$ is the density of the chondrule.

In practice, in order for the chondrules to have the properties
observed, they must cool on a much longer timescale than this,
i.e. $t_{\rm cool} \gg t_{\rm rad}$ (Scott 2007). Estimates of
cooling times vary greatly. Levi \& Araki (1989) suggest $t_{\rm cool}
\sim 2 \times 10^3$ to $2 \times 10^4$~s. Hutchinson (2004) agrees
with this at the short end, but indicates that some cooling times must
be much longer, in the range $t_{\rm cool} \sim 2 \times 10^5$ to $3
\times 10^6$ s. From the detailed discussion given by Desch \&
Connolly (2002) it is evident that the cooling histories of chondrules
are varied and complicated. Scott et al. (1996) argue that the disc
must have been `a maelstrom where temperatures fluctuated through
$\sim 1000$~K and solids experienced ``multiple cycles of melting,
evaporation, recondensation, crystallization and aggregation'' '.

Desch \& Connolly (2002) give characteristic cooling times in the
range of hours to days. This is in line with the recent review by
Connolly et al. (2006), who quote the majority of chondules as having
a cooling time of $\sim 10^5$~s. Thus we adopt a typical cooling time
for chondrules of
\begin{equation}
t_{\rm cool} = 10^5 \, {\rm s}.
\end{equation}

The fact that $t_{\rm cool} \gg t_{\rm rad}$ has implications for the
energetics of the formation process. Not only must the chondrules be
heated to melting temperature, but they must also be kept near that
temperature for a prolonged period. This implies that not just the
proto--chondrules but also the material surrounding them must be
heated to around 2000 K (Wasson, 1996).
In addition, a sufficient volume of
surrounding material must be heated so that its optical thickness
implies a cooling time of hours to days.  The specific energy required
to heat the surrounding gas to a temperature of $T = 2000$ K is given
approximately by
\begin{equation}
E_{\rm requ} \approx \frac{\frac{3}{2} kT}{2 m_H} = 1.2 \times 10^{11}
\left( \frac{T}{2000 {\rm K}} \right) \, {\rm erg \, g}^{-1},
\end{equation}
where we have assumed that the typical gas particle is a hydrogen
molecule, and have ignored molecular dissociation. This implies,
again assuming that on average each element of chondritic material is
subject to one melting event, that a fraction $f$ of around
\begin{equation}
\label{fraction}
f \sim \frac{E_{\rm requ}}{E_{\rm kin}} \approx 0.08
\end{equation}
of the available accretion energy must be used to provide the heating
process for chondrule formation.

\section{Shock heating}

Desch \& Connolly (2002) have proposed a detailed model for the
formation of chondrules in terms of the thermal processing of
particles in shocks in the accretion disc. Their model takes account
of the properties of the shock, including gas--drag heating of
particles which are not slowed instantly in the gas shock, radiative
processes including the disassociation and recombination of H$_2$, and
particle evaporation. Their model can account for the thermal history
of chondrules. Their canonical model requires a shock velocity of
\begin{equation}
V_s = 7 \times 10^5 {\rm cm \, s}^{-1}.
\end{equation}
For their assumed ambient disc temperature of $T = 300$ K and so sound
speed (assuming a gas of molecular hydrogen) of $c_s = 1.5 \times
10^5$ cm s$^{-1}$, this corresponds to a Mach number of around ${\cal
  M} \sim 5$.

However, while this model can successfully account for chondrule
formation in the disc, the proposed source of the shocks within the
disc does not sit easily with our current understanding of
protostellar disc evolution.  Desch \& Connolly (2002) propose that
the shocks are caused by gravitational instabilities within the
disc. But this proposal is at odds with their disc properties, which
are taken from a disc model by Bell et al. (1997) with an accretion
rate of $\dot{M} = 10^{-8} M_\odot$ yr$^{-1}$ and a viscous parameter
of $\alpha = 10^{-4}$.  The disc density is assumed to be $\rho =
10^{-9}$ g cm$^{-3}$, which for a disc semi-thickness $H =
(c_s/V_\phi) R \approx 4 \times 10^{12}$ cm gives a disc surface
density of $\Sigma = 8000$ g cm$^{-2}$. This gives rise to a strongly
unstable disc with a Toomre parameter (Toomre 1964) of
\begin{equation}
Q = \frac{\Omega c_s}{\pi G \Sigma} \approx 3.4 > 1.
\end{equation} 
Such strong gravitational instability is  indeed required if this process is
to give rise to shocks with velocities $V_s \approx 0.4 V_\phi$. It
is, however, hard to envisage how all the material in such an strongly
unstable disc would manage to avoid being strongly shocked a large
number of times.

Such a disc has a mass at radii around 3 AU (say, at radii 2 -- 4 AU)
of
\begin{equation}
M_{\rm disc}(3 {\rm AU}) \approx \pi R^2 \Sigma \approx 0.025 M_\odot,
\end{equation}
Such a violently unstable disc is likely either to fragment into
predominantly gaseous bodies, gathering any chondrules with them, or,
if not, give rise to strong gravitational torquing and so a high
accretion rate.  The accretion rate expected for a disc which is so
gravitationally unstable is the same as one for which the viscous
parameter has $\alpha \approx 0.06$ (Lodato \& Rice, 2004; Rice,
Lodato \& Armitage, 2005) which would give (Pringle 1981)
\begin{equation}
\dot{M} = 3 \pi \alpha c_s H \Sigma \approx  4 \times 10^{-5} M_\odot
{\rm yr}^{-1}.
\end{equation}

Such a high accretion rate would give a midplane disc temperature of
$\sim 2000$~K (Bell et al., 1997) and contradicts the original
assumption that $\dot{M} = 10^{-8} M_\odot$ yr$^{-1}$. It would give a
local disc lifetime of around $M_{\rm disc}/\dot{M} \approx 600$ yr.

Similar considerations of accretion rates and lifetimes apply to the
more detailed models presented by Boss \& Durisen (2005; see also
Boss, 2007).  These authors stress that early times, when infall onto
the disc was high, the disc is most likely massive enough for
self-gravity to play a major role in angular momentum redistribution
(see, for example, Lin \& Pringle, 1990). They also note, however,
that the local, non-axisymmetric instabilities driven by self-gravity
occur at, or around, co-rotation (see also the discussion in Cossins
et al., 2009). This implies that in general, throughout the bulk of
the disc, the relative velocity between disc gas and the (unstable,
but transient) spiral pattern is small. Thus, in general, the
short-lived, localised, transient spirals which are the typical
outcomes of self-gravitational disc instabilities, are not able to
drive strong shocks in the gas. In this context, in the numerical
simulations of Boss \& Durisen (2005) it is only close to (within a
few grid cells of) the inner grid boundary that they are able to
achieve a shock pattern which is neither close to co-rotation, nor
strongly trailing.

Thus while we cannot rule out the possibility that a massive disc and
self-gravitational instabilities are able to provide the necessary
shocks, we now consider what we regard as more likely properties for
the disc in the region of a few AU at the time of chondrule formation.

\section{Disc Properties}

From timings deduced from isotopic data, Russell et al. (2006)
conclude that chondrule formation lasted for around 3 -- 5 Myr. This
is also the timescale on which young solar--type stars appear to lose
their protoplanetary discs (Hernandez et al. 2009). Thus it seems
reasonable to conclude that chondrules were formed during the final
stages of the protoplanetary disc when the accretion rates were low
($\dot{M} \le 10^{-7} M_\odot$ yr$^{-1}$, Ciesla \& Charnley, 2006).

Models of protostellar discs at low steady accretion rates are given
by Bell et al (1997). At the accretion rates we consider, it is
likely that heating from the central star provides a non--negligible,
and perhaps dominant, contribution in the surface layers, which may
cause some changes on the midplane (e.g. Chiang et al., 2001). There
is also the possibility that such discs contain `dead zones' (Balbus
\& Hawley 2000; Terquem, 2008) where the MRI is unable to operate
because of high magnetic diffusivity caused by low ionization. However
for the canonical disc parameters we use (see below) Matsumura \&
Pudritz (2003) argue that the MRI should be fully operational. This is
in line with our argument above that the full accretion energy
must be accessed to power chondrule formation.

At the time of chondrule formation we take the disc mass as a few
times the minimum mass solar nebula (say, $ \sim 0.06 M_\odot$) and its
age as $\sim 3$ Myr. An accretion rate of order $\dot{M} \sim 3 \times
10^{-8} M_\odot$ yr$^{-1}$ is appropriate. Given this, the models of
Bell et al (1997) suggest the following parameters for the disc in the
region of $R \approx 3$ AU. The disc (midplane) density is
\begin{equation}
\rho_{\rm disc}= 10^{-10} {\rm g \, cm}^{-3},
\end{equation}
and the midplane temperature is
\begin{equation}
T_{\rm disc} = 200 {\rm K},
\end{equation}
with disc thickness
\begin{equation}
H = 0.05 R = 2.2 \times 10^{12} {\rm cm}
\end{equation}
and viscosity parameter
\begin{equation}
\alpha = 0.01.
\end{equation}

In this model most of the disc mass still resides at radii 10 -- 20
AU, and it is these radii which control the evolution timescale of the
disc, and so the accretion rate. These parameters are in line with the
fully ionized models of Terquem (2008) and with the disc properties
assumed by Ida \& Lin (2004) in their modelling of planet core
formation. For these parameters the disc sound speed is
\begin{equation}
c_s = 1.2 \times 10^5 \, {\rm cm \, s}^{-1},
\end{equation}
and the gas pressure is
\begin{equation}
p_g = 0.83 \, {\rm dyne \, cm}^{-2}.
\end{equation}

\section{Magnetic fields}

Remnant magnetic fields found in chondritic material imply cooling
through the Curie temperature of $\sim 600$ K in the presence of a magnetic
field. This offers clear evidence that chondrule formation took
place in regions with magnetic fields in the range $B \sim 1 - 10$ G
(Levy \& Sonett, 1978; Levy \& Araki, 1989). This is close to the
maximum (equipartition) field strength $B$ of a magnetic flux tube in
an accretion disc -- equating the magnetic pressure $B^2/8 \pi$ with the
local gas pressure $p_g$ for the disc parameters given above gives
\begin{equation}
\label{Beq}
B_{\rm eq} = 4.6 \, {\rm G}.
\end{equation}
We note that to achieve the strength of viscosity ($\alpha =
0.01$) assumed above, we only require an average magnetic field
$\langle B \rangle$ within the disc given by (Shakura \& Sunyaev,
1973)
\begin{equation}
\alpha \approx \frac{\langle B \rangle^2}{8 \pi p_g},
\end{equation}
i.e.
\begin{equation}
  \langle B \rangle \approx 0.5 \left( \frac{\alpha}{0.01} \right)^{1/2}
  \, {\rm G}.
\end{equation}
This suggests that chondrule formation could have taken place in a
disc in which a dynamo mechanism driven by the magneto-rotational
instability was operating.

\section{The dynamo process and chondrule formation}

We have seen that chondrule formation requires the heating of each
element of the disc (gas plus proto--chondrules) up to a temperature
of around $T \approx 2000 $ K $ \approx 10 T_{\rm disc}$, for a time
$\sim t_{\rm cool} = 10^5$~s, on average once. This must happen
during the time that this material is at a radius of order 3 AU. Once
the gas is at this temperature it becomes over--pressured relative to
its surroundings, and so will tend to expand, and therefore cool
adiabatically, at its now elevated sound speed of $c_s^\ast \approx
3.8 \times 10^5$ cm s$^{-1}$. For this adiabatic cooling to take
longer than $t_{\rm cool}$ we require the size of the heated
region $h$ to be such that
\begin{equation}
h \ge c_s^\ast \times t_{\rm cool} \approx 3.8 \times 10^{10} {\rm cm}.
\end{equation}

We have also noted (Deschl \& Connolly, 2002) that a way of achieving
such heating is through shocks within the disc, with shock velocities
$V_s \sim 7 \times 10^5$ cm s$^{-1} \sim 6 \, c_s$. However, although
the standard shearing--box models of MRI--driven dynamos in accretion
discs (see Davis, Stone \& Pessah, 2009) can
produce values of $\alpha$ of order 0.01, they predict velocities within
the disc which are for the most part strongly subsonic. In this picture,
chondrule formation would thus remain unexplained.

However, there are indications that such models do not give a complete
description of dynamo processes in accretion discs (cf King, Pringle
\& Livio, 2007). One problem is that these models use the MHD
approximation with a standard (isotropic) magnetic diffusivity $\eta$
and standard (isotropic) Navier--Stokes viscosity $\nu$. Fromang et
al. (2007) demonstrate that the results obtained depend on the assumed
numerical parameters: the Reynolds number ${\cal R} = c_sH/\nu$, the
magnetic Reynolds number ${\cal R}_m = c_sH\eta$ and, critically, on
the ratio of the two, the magnetic Prandtl number ${\cal P}_m = {\cal
  R}_m/{\cal R} = \nu/\eta$. There are also other physical processes
which can provide complications such as the Hall effect (Balbus \&
Hawley, 2000), and anisotropic transport processes (e.g. Dong \&
Stone, 2009). Schekochihin and co-workers (e.g. Schekochihin et al.,
2004, 2005) have also argued that the structure of a magnetic field in
a turbulent flow at such high Reynolds numbers can depend critically
on the Prandtl number, and that this has implications for dynamos
under such conditions. In addition, Heitsch et al. (2008) and Zweibel
\& Heitsch (2008) have discussed the implications for magnetic growth
and structure in turbulent media where ambipolar diffusion plays a
significant role.

\subsection{An alternative dynamo picture}

In the light of the above, we speculate how an accretion disc dynamo
might produce the intermittent high--energy events which appear to be
demanded by chondrule formation. As an example we consider a new model
for a dynamo in a turbulent medium suggested by Baggaley et
al. (2009a,b).  The extent to which such a model is applicable to
cooler accretion discs, such as the solar nebula, is somewhat
uncertain, and we introduce it here because it is able to illustrate
the kind of properties with regard to energy release that we are
looking for to facilitate chondrule formation.  In this model the
magnetic field is mainly confined to
thin flux ropes which are advected by the flow. Magnetic dissipation
only occurs via reconnections of the flux ropes, and so the magnetic
dissipation is highly localised. This model can be viewed as an
implementation of the limiting regime of infinitely large
magnetic Reynolds number: magnetic dissipation can be safely neglected
at all scales, but plays a crucial role through reconnection of field
lines in permitting rearrangement of the field topology. Such
rearrangements of the field topology result in conversion of magnetic
energy into kinetic energy of the fluid, and thence to dissipation as
heat. This contrasts to the usual models in which magnetic diffusivity
converts magnetic energy to heat directly.

With this in mind, the picture we propose is that the magnetic field
in the disc can be thought of as a collection of loops of flux
ropes. These loops are continually stretched by the azimuthal shear
flow. The stretching increases the magnetic energy associated with the
loop and at constant total (gas plus magnetic) pressure decreases the
mass density along the field line, and so increases the Alfv\'en speed
along the loop.  To maintain an equilibrium distribution of loop
properties, the loops must also continually undergo reconnection
events. The crucial property of these events is that while each
reconnection event in itself releases a negligible amount of energy,
it does release the magnetic field, which is then able to reconfigure
itself (at the Alfv\'en speed) and in so doing, heat the gas. The
picture here thus differs fundamentally from that presented by Sonett
(1979) and Levy \& Araki (1989). They considered magnetic reconnection
in low density regions far from the disc plane (in the disc corona)
and took only the magnetic energy dissipated by the reconnection
events into account. We are assuming here that most of the accretion
energy is released in the bulk of the disc. We should note, however,
that some authors have suggested that a substantial fraction of the
accretion energy might be released in such low density regions (Tout
\& Pringle, 1990; Uzdensky \& Goodman, 2008), arguing that magnetic
buoyancy can advect energy efficiently away from the disc plane. For
our picture to work, however, we require that a significant amount of
reconnection occurs close to the disc plane, so that the bulk of the
energy release occurs there and disc gas, along with the chondrule,
can be efficiently shock heated.

What we therefore require for chondrule formation is that {\it some}
reconnection events occur in regions with sufficiently high Alfv\'en
speeds that shock velocities of $V_s \approx 6 c_s$ can be
generated. For this to occur we require that a minority of flux tubes
have field strengths of order $B_{\rm eq}$ and loop mass densities
around 40 times lower than the mean disc density before reconnection
occurs. Unfortunately the `fluctuation dynamo' model of Baggaley et al
(2009a, b) is currently computed only in an incompressible medium. Further
consideration of the model will be required before it is possible to
establish whether or not such high--energy reconnection events are
likely in an accretion disc.

\subsection{Implications for magnetic field structure and internal 
disc properties}

In this picture we suppose that from time to time a reconnection event
occurs which causes a sufficiently large and rapid adjustment of field
topology that a region of disc gas of size $h$ is subject to shock
heating, with velocity $V_s \approx 7$ km s$^{-1}$. We need this
heated region to stay sufficiently hot for a time $t_{\rm cool}
\approx 10^5$~s. We can use this information to deduce requirements
for the properties of the magnetic loops.

\subsubsection{Size of heated regions}

We expect the cooling timescale $\tau$ for a region of size $h$
and temperature $T$ to be given by
\begin{equation}
\tau \sim \frac{\rm heat \, content}{\rm heat \, loss \, rate}.
\end{equation}
We expect that the heat content is simply
\begin{equation}
{\rm heat \, content} \propto \rho T h^3,
\end{equation}
where $\rho$ is the gas density.
If the region is optically thick, so that heat loss is mainly
by radiative transfer, then we expect
\begin{equation}
{\rm heat \, loss \, rate} \propto \frac{T^4}{ \rho \kappa h} \times h^2,
\end{equation}
where $\kappa$ is the opacity.
Thus we expect
\begin{equation}
\label{tau}
\tau \propto \frac{\rho^2 \kappa h^2}{T^3}.
\end{equation}

For an accretion disc in thermal equilibrium, the relevant
lengthscale is the disc thickness $H$, and the cooling timescale is
the thermal timescale $t_{\rm th}$,  given by (Pringle 1981)
\begin{equation}
t_{\rm th} \approx \frac{1}{\alpha \Omega} \approx 2.6 \times 10^9
\left( \frac{\alpha}{0.01} \right)^{-1} \, {\rm s}.
\end{equation}

Using this we can obtain a very rough estimate of the size required
for the shock--heated regions. To keep things simple we assume that
the opacity $\kappa \approx$ const. (e.g. Bell et al., 1997), although
if some chondritic material is vaporised and/or H$_2$ is significantly
dissociated, this might not be the case. Then from the computations of
Deschl \& Connolly (2002) we note that once the gas has relaxed behind
the shock, the temperature is $T \approx 1000$ K, and so a factor of
$\sim 50$ above its pre--shock value, and similarly that the density
$\rho$ has increased by a factor of order $\sim 10$. Thus using
Equation~\ref{tau} we find
\begin{equation}
\frac{h^2}{H^2} \sim \frac{t_{\rm cool}}{t_{\rm th}} \times (50)^3
  \times (10^{-2}),
\end{equation}
which gives
\begin{equation}
h \sim 0.2 H \sim 4 \times 10^{11} \, {\rm cm}.
\end{equation}

\subsubsection{Heating events and loops}

We have seen that a single heating event is likely to have a size of
around $h \sim 4 \times 10^{11}$ cm. Evidently this must be roughly
the size of the reconnecting loops. This means that the rate per unit
volume at which energy is dissipated {\it locally} by the shock is
approximately
\begin{equation}
q_{\rm shock} \sim \rho V_s^2 \times \left( \frac{V_s}{h} \right).
\end{equation}

We can contrast this with an accretion disc in local thermal
equilibrium, where the much lower {\it average} volume rate at which
energy is dissipated within the disc is approximately
\begin{equation}
q^+ = \frac{\rho c_s^2}{t_{\rm th}}.
\end{equation}

We can now use this to estimate the probability $p$ that a random disc
element is hot at any one time. The average rate at which shocks heat
the disc is
\begin{equation}
\langle q_s \rangle \approx p \times q_{\rm shock}. 
\end{equation}
We have argued earlier (eqn~\ref{fraction}) that to heat on
average each element of disc gas up to a temperature of around $T
\approx 2000$ K once every viscous timescale requires about $f \approx
0.08$ of the available accretion energy. Thus we require that
\begin{equation}
\langle q_s \rangle \approx f \times q^+,
\end{equation}
and hence that
\begin{equation}
p \approx f \, \left( \frac{c_s^2}{V^2} \right) \frac{h/V_s}{t_{\rm
    th}} \approx 3.3 \times 10^{-7}.
\end{equation}

We have argued that each chondrule heating event should take place on a
scale of $\sim h \approx 4 \times 10^{11}$ cm. If we divide the disc
locally (say 2 -- 4 AU) into $N_{\rm box}$ boxes, each of size $\sim
h^3$ then
\begin{equation}
N_{\rm box} \sim \pi R^2.2H/h^3 \sim 1.4 \times 10^5.
\end{equation}

Therefore the number of these boxes which is `hot' at any one time is 
\begin{equation}
N_{\rm hot} \approx p \times N_{\rm box} \approx 4.6 \times 10^{-2}.
\end{equation}
Since the orbital period at this radius is $P = 1.6 \times 10^8$ s,
each box is therefore heated for a fraction $f_{\rm hot} \approx
(h/V_s)/P \approx 3.6 \times 10^{-3}$ of an orbit. Similarly the
average time between these large flare events is given by $t_{\rm
  flare} \sim P/ N_{\rm hot} \approx 100$ yr.

Assuming that the reconnection rate for loops is roughly the same
as the rate at which they are stretched by the disc shear flow, so that
on average each loop reconnects once per orbit, then the number of
reconnections per orbit which give rise to major flares, and so chondrule
heating, is
\begin{equation}
N_{\rm requ} \sim N_{\rm hot} / f_{\rm hot} \sim 13.
\end{equation}

\subsubsection{Loop strength distribution and relation to the dynamo}

We have argued that only about $f \approx 0.08$
(eqn~\ref{fraction}) of the energy is released in heating events
which are powerful enough to cause chondrule formation. If we
make the simplifying assumption that magnetic loops are all
approximately the same physical size ($\sim h$) but have varying
magnetic field strength, then this implies that most loops have only
weak field strengths. This ties in with the argument
(eqn~\ref{Beq}) that to obtain a viscosity parameter of $\alpha
\approx 0.01$ we only require an average disc field of order $\langle
B \rangle \approx 0.5$ G.

If all the loops are the same size, then the total number of loops
would be
\begin{equation}
N_{\rm loops} \sim N_{\rm box} \sim 1.4 \times 10^5.
\end{equation} 
If, in a steady state, each loop reconnects once per dynamical time
($\approx$ orbital period) then over this time there are a total of
$N_{\rm loops} \sim 1.4 \times 10^5$ reconnections, of which $N_{\rm
  requ} \sim 13$ give enhanced heating, and so chondrule
formation. These strong flares provide a fraction $f \approx 0.08$ of
the total power of the disc, and so the energy released in a weak
reconnection event is $\sim N_{\rm requ}/f N_{\rm loops} \sim 1.3
\times 10^{-3}$ times that released in a strong one. The frequency
distribution $f(E)$ for events which release an energy $E$ (cf.
Charbonneau et al., 2001) is usually assumed to take a power law form
\begin{equation}
f(E) \propto E^{-s}, \: s>0.
\end{equation}
Then our reasoning above implies that $s \approx 2.3$.  This is
similar to what is found for the energy release in solar flares, for
which $s$ is generally in the range $1.4 \la s \la 2.6$ (Charbonneau
et al., 2001). Note that $s=2$ implies equal energy release at all
scales, so that $s = 2.3$ implies that energy release occurs mainly
(but marginally so) at small scales. For their incompressible loop
dynamo model, Baggaley et al. (2009b) find $s \approx 3$, so that small
scales strongly dominate.

\section{Conclusions}

We have argued that the need for strong intermittent heating during
the formation of chondrules in the protosolar nebula places important
constraints on the way gravitational energy is released in accretion
discs. Following Desch \& Connoly (2002) we suggest that this requires
shock heating,and propose that the shocks occur as magnetic loops in
the disc dynamo are released by reconnection events. In contrast,
current shearing--box simulations of MRI--driven accretion disc
dynamos predict largely subsonic velocity fields, and so leave the
formation of chondrules unexplained.

We have proposed that accretion disc dynamos may be much more
inhomogeneous, in space and in time, than current simulations
indicate. It would not be surprising if the input physics to current
numerical simulation was incomplete, and we have briefly discussed
reasons why this might be so. We speculate that a better understanding
of accretion disc energy release may require consideration of more
complicated physical processes than are currently implemented in
numerical codes.

\section{Acknowledgments}

We thank Mark Hurn for valuable help in finding references. We thank
Alan Boss for useful comments and for sending us a copy of his
review. We thank Andrew Baggaley and Jim Stone for stimulating
correspondence. We acknowledge support from the Isaac Newton Institute
programme `Dynamics of Discs and Planets'.

\label{lastpage}

\end{document}